\def\year{2017}\relax
\documentclass[letterpaper]{article}
\usepackage{aaai17}
\usepackage{times}
\usepackage{helvet}
\usepackage{courier}
\usepackage{multicol}
\usepackage{graphicx}
\usepackage{caption}
\usepackage{subcaption}
\usepackage{algorithm}
\usepackage{algorithmic}
\usepackage{amsmath}
\begin{document}
\title{Quality Enhancement by Weighted Rank Aggregation of Crowd Opinion}
\author{Sujoy Chatterjee$^1$, Anirban Mukhopadhyay$^1$ and Malay Bhattacharyya$^2$\\
$^1$Department of Computer Science and Engineering, University of Kalyani, Nadia -- 741235, India\\
E-mail: \{sujoy, anirban\}@klyuniv.ac.in\\
$^2$Department of Information Technology, Indian Institute of Engineering Science and Technology, Shibpur\\ Howrah -- 711103, India\\
E-mail: malaybhattacharyya@it.iiests.ac.in\\
}
\maketitle

\begin{abstract}
Expertise of annotators has a major role in crowdsourcing based opinion aggregation models. In such frameworks, accuracy and biasness of annotators are occasionally taken as important features and based on them priority of the annotators are assigned. But instead of relying on a single feature, multiple features can be considered and separate rankings can be produced to judge the annotators properly. Finally, the aggregation of those rankings with perfect weightage can be done with an aim to produce better ground truth prediction. Here, we propose a novel weighted rank aggregation method and its efficacy with respect to other existing approaches is shown on artificial dataset. The effectiveness of weighted rank aggregation to enhance quality prediction is also shown by applying it on an Amazon Mechanical Turk (AMT) dataset.
\end{abstract}

\section{Introduction}
Ranking of various objects is useful in numerous real-life problems \cite{Dwork2001,DeConde2006}. There are several problems that consider conflicting settings that require to find out better solution from multiple alternatives. In these models, the quality of a solution plays a vital role in predicting the final judgment. Basically different quality metric criteria can be taken to judge the quality of a solution. The reason is that based on a single quality metric the competence level of a solution can not be estimated properly. Therefore, aggregation of those rankings is very much needed to produce final ranking. Rank aggregation \cite{DeConde2006,Dwork2001,Stuart03,Kolde2012,Chen:2013:PRA:2433396.2433420} has been studied in depth over the years for combining multiple alternatives so as to apply in diverse research applications including bioinformatics, crowdsourcing, etc.

In some crowdsourcing studies \cite{Hovy2013,Ipeirotis2010,Liu2013,Nguyen16-hcomp,ChatterjeeS2017}, there is a need to find consensus from multiple crowdsourced opinions. Straightforward aggregation policy can not be applied as there is a possibility of inclusion of spamming in these opinions. Now, it is observed that in many problems, relying on multiple features can be effective to achieve accurate prediction. In this paper, we introduce a novel weighted rank aggregation method considering position based score, while ranking is done depending on various features like specificity, accuracy, sensitivity, etc. To show how this aggregation method is itself good, it is compared with state-of-the-art rank aggregation approaches over a few artificial datasets. Moreover, the quality of ground truth prediction is also tested over a real-life AMT crowdsourcing dataset.

\section{Problem Formulation}
We consider a set of rankings $R$ = $\{R_1, R_2, \ldots, R_m\}$ and a set of weights $\{w_1, w_2, \ldots, w_m\}$ for each of the ranking. Suppose $c$ is a distance function to compute the closeness between any two rankings. Now, the objective is to find out the ranking that have minimum distance $\delta^* $ from each of the individual ranking $R_i$ in the ranking list $R$. This problem can be mathematically expressed as follows

\begin{eqnarray}
\delta^* = \arg min {\sum_{i} w_i*c(R_i,\delta)}.
\end{eqnarray}

The distance $c$ can be any measure, i.e., Kendall's tau distance or Spearman's footrule distance. Here, $w_i$ denotes the weight of the corresponding ranking $R_i$.

\section{Proposed Model}
Suppose, $R_1$ = \{1,2,4,3,5\} and $R_2$ = \{2,1,3,4,5\} are two different orderings of 5 objects. If such multiple ranking exist then it is challenging to produce a consensus decision. We now introduce a few basic terms that might be useful for a better understanding of the proposed model.

{\bf Gain Score:} The Gain score obtained by an object is due to its precedence of position with respect to another object in a particular ranking. So, if there are $m$ objects, then the gain score obtained by the $1^{st}$ object in the list is due to its precedence over $(m-1)$ objects.
In general, the gain score $\mathcal{GS}_k$ for any object with position $k$ is expressed by $(m-k)(m-k-1)/2$. %In this context, the starting value of position is considered as 0 and therefore the object that is in first position (within a input ranking) with $k=0$ has maximum gain score.

{\bf Penalty score:} The penalty score incurred by any object is due to its position lagging behind in a particular ranking. The penalty score $\mathcal{PS}_k$ received by an object with position $k$ is $k(k+1)/2$.

{\bf Overall score:} The position based overall score $\mathcal{SC}_k$ for an object with $k^{th}$ position is calculated by subtracting penalty score $\mathcal{PS}_k$ from gain score $\mathcal{GS}_k$ and it is expressed as $\frac{m^2 - 2m*k - m}{2}$.

Now, we illustrate the overall procedure to find out an aggregated ranking from a list of input rankings.

\begin{itemize}
\item {\bf Step 1:} Initially, a similarity matrix between different rankings is computed using the weighted Spearman's footrule similarity measure.

\item {\bf Step 2:} From the similarity matrix, the two most similar rankings are chosen and they are merged in order to get more accurate ranking. The merging of two rankings is done by using the corresponding weight of these rankings. This merging technique takes into account the goodness of the parent rankings.

Now if the weight of ranking $R_1$ is $w(R_1)$ and weight of ranking $R_2$ is $w(R_2)$, then according to the formula, the merge score $\mathcal{MS}^k$ of the object $k$ is calculated as follows.

\begin{equation}\label{eq:merge_eqation}
\mathcal {MS}^k  = \frac{w(R_1)*a + w(R_2)*b}{w(R_1) + w(R_2)},
\end{equation}

where $a$, $b$ are the scores of the same object based on the positions in two different rankings. In this way, the merged scores of all the objects are computed. Finally, the ranking can be obtained after sorting all the merged scores. If multiple rankings contain similar values then all the possible merging are computed and suitable ranking with minimum distance to all input ranking is chosen.

\item {\bf Step 3:} In this process, the weight of the new ranking is also computed. To compute the weight of new ranking, past experience (goodness of the parent rankings) as well as the present fitness (based on the closeness of new ranking with all the input rankings) are considered.

\item {\bf Step 4:} The steps 1-3 are repeated until a single aggregated ranking is reached.
\end{itemize}

\section{Experimental Results}
\begin{figure}[!t]
\centering
\includegraphics[width=5.7cm,height=3cm]{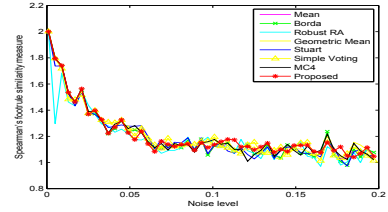}
\caption{Comparative values of Spearman's footrule similarity measure between the aggregated ranking and all of the input rankings with the increment of noise.}
\label{Fig:plot}
\end{figure}

We have artificially generated several datasets with different dimensions by adding Gaussian noise to each input rank list in a step-wise manner. Inclusion of noise means position of a few bits are altered from the original ranking. When the noise is maximum, most of the bits are altered. For the dataset, the algorithm is applied for 50 iterations and in each iteration the Gaussian noise is incremented slowly by varying the standard deviation. Now, in each iteration (i.e., for any particular step), the rank aggregation algorithm is applied on the input ranking list and the weighted similarity values (Spearman's footrule distance) between the resultant aggregated ranking and the input rankings are computed (see Fig.~\ref{Fig:plot}). Finally, to compute the area under the curve (AUC) trapezoidal function is applied. The AUC values for different rank aggregation algorithms and for the proposed approach are given in Table~\ref{table:result_aggregation1}.

\begin{table}[!t]
\small
\renewcommand{\arraystretch}{1.0}%
\centering
\caption{The AUC values obtained for the proposed approach and other existing approaches for different instances of a dataset with dimension $20 \times 30$. Here, the standard deviation is varied by 0.004 (Instance 1), 0.01 (Instance 2) and 0.02 (Instance 3) in each step. The best scores over a column (i.e., for a particular instance) are shown in bold.}
\begin{tabular}{|c|c|c|c|c|}
\hline
{\bf Algorithm}     & {\bf Instance 1} & {\bf Instance 2}  & {\bf Instance 3}  \\\hline
MC4           & 0.2360       & 0.3394      & 1.0359   \\\hline
Robust RA 	  & 0.2301       & 0.3227      & 1.0124  \\\hline
Mean          & 0.2342       & 0.3378      & 1.0273  \\\hline
Geometric     & 0.2356       & 0.3385      & 1.0344   \\\hline
Stuart        & 0.2345       & 0.3364      & 1.0314  \\\hline
Simple Voting & 0.2363       & 0.3398      &  1.0325   \\\hline
Borda         & 0.2344       & 0.3379      & 1.0296  \\\hline
Proposed      & {\bf 0.2367} & {\bf 0.3416}   & {\bf 1.0406}  \\\hline
\end{tabular}%}
\label{table:result_aggregation1}
\end{table}

To investigate the performance of the proposed method in crowdsourcing domain, the method is applied on real-life RTE dataset \cite{Sheshadri2013}. In this experiment, we have used majority voting as the baseline depending on which the accuracy, specificity, sensitivity and recall of the annotators are computed. Finally, the consensus ranking of them are calculated by aggregating those list of ranks and the final aggregated opinion is predicted. Superiority of the scheme in terms of accuracy is reported in Table~\ref{table:crowd_aggregation}.

\begin{table}[!t]
\small
\renewcommand{\arraystretch}{1.0}%
\centering
\caption{Accuracy values obtained for different opinion aggregation approaches by applying on the RTE dataset. The best accuracy scores over a column  are shown in bold.}
%\resizebox{\textwidth}{!}{%
\begin{tabular}{|c|c|}
\hline
{\bf Algorithm}  & {\bf RTE Dataset} \\\hline
Majority Voting  & 89.88\%       \\\hline
MACE  \cite{Hovy2013}	& 93.00\%   \\\hline
Raykar \cite{Rayker2011} & 93.00\%    \\\hline
GLAD \cite{Whitehill2009} & 78.7\%  \\\hline
DS \cite{Dawid1979} & 82.2\%     \\\hline
Proposed  & {\bf 93.37\%}     \\\hline
\end{tabular}%}
\label{table:crowd_aggregation}
\end{table}

\section{Conclusion}
In this paper, we show the utility of weighted rank aggregation in the field of crowd based judgment analysis problem. To investigate how good the proposed method is, with respect to other methods, the applications has been done on a real-life AMT dataset. The model can be made more robust if the crowd workers' self-reported confidence scores are taken as input while collecting the annotation scores.

\section{Acknowledgment}
The work of Malay Bhattacharyya is supported by the Visvesvaraya Young Faculty Research Fellowship 2015-16 of MeitY, Government of India. All the authors would like to thank the crowd contributors involved in this work.

%\clearpage
%\subsection{References}
\bibliographystyle{aaai}
\bibliography{Reference}

\begin{thebibliography}{}

\bibitem[\protect\citeauthoryear{Chatterjee and
  Bhattacharyya}{2017}]{ChatterjeeS2017}
Chatterjee, S., and Bhattacharyya, M.
\newblock 2017.
\newblock Judgment analysis of crowdsourced opinions using biclustering.
\newblock {\em Information Sciences} 375:138--154.

\bibitem[\protect\citeauthoryear{Chen \bgroup et al\mbox.\egroup
  }{2013}]{Chen:2013:PRA:2433396.2433420}
Chen, X.; Bennett, P.~N.; Collins-Thompson, K.; and Horvitz, E.
\newblock 2013.
\newblock Pairwise ranking aggregation in a crowdsourced setting.
\newblock In {\em Proceedings of the Sixth ACM International Conference on Web
  Search and Data Mining}, WSDM '13,  193--202.
\newblock New York, NY, USA: ACM.

\bibitem[\protect\citeauthoryear{Dawid and Skene}{1979}]{Dawid1979}
Dawid, A.~P., and Skene, A.~M.
\newblock 1979.
\newblock {Maximum likelihood estimation of observer error-rates using the EM
  algorithm.}
\newblock {\em Applied Statistics} 28(1):20--28.

\bibitem[\protect\citeauthoryear{DeConde \bgroup et al\mbox.\egroup
  }{2006}]{DeConde2006}
DeConde, R.; Hawley, S.; Falcon, S.; Clegg, N.; Knudsen, B.; and Etzioni, R.
\newblock 2006.
\newblock {Combining Results of Microarray Experiments: A Rank Aggregation
  Approach.}
\newblock {\em Statistical Applications in Genetics Molecular Biology}
  5(1):1--17.

\bibitem[\protect\citeauthoryear{Dwork \bgroup et al\mbox.\egroup
  }{2001}]{Dwork2001}
Dwork, C.; Kumar, R.; Naor, M.; and Sivakumar., D.
\newblock 2001.
\newblock Rank aggregation methods for the web.
\newblock In {\em Proceedings of the Tenth International World Wide Web
  Conference, ACM},  613–--622.

\bibitem[\protect\citeauthoryear{Hovy \bgroup et al\mbox.\egroup
  }{2013}]{Hovy2013}
Hovy, D.; Kirkpatrick, T.~B.; Vaswani, A.; and Hovy, E.
\newblock 2013.
\newblock Learning whom to trust with mace.
\newblock In {\em Proceedings of the North American Chapter of the Association
  for Computational Linguistics -- Human Language Technologies (NAACL HLT)},
  1120--1130.

\bibitem[\protect\citeauthoryear{Ipeirotis}{2010}]{Ipeirotis2010}
Ipeirotis, P.
\newblock 2010.
\newblock Analyzing the amazon mechanical turk marketplace.
\newblock {\em ACM XRDS} 17(2):16--21.

\bibitem[\protect\citeauthoryear{Kolde \bgroup et al\mbox.\egroup
  }{2012}]{Kolde2012}
Kolde, R.; Laur, S.; Adler, P.; and Vilo, J.
\newblock 2012.
\newblock {Robust rank aggregation for gene list integration and
  meta-analysis.}
\newblock {\em Bioinformatics} 28(4):573--580.

\bibitem[\protect\citeauthoryear{Liu, Peng, and Ihler}{2013}]{Liu2013}
Liu, Q.; Peng, J.; and Ihler, A.
\newblock 2013.
\newblock Report of crowdscale shared task challenge 2013.
\newblock In {\em Proceedings of the Crowdscale Shared Task Challenge}.

\bibitem[\protect\citeauthoryear{Nguyen \bgroup et al\mbox.\egroup
  }{2016}]{Nguyen16-hcomp}
Nguyen, A.~T.; Halpern, M.; Wallace, B.~C.; and Lease, M.
\newblock 2016.
\newblock {Probabilistic Modeling for Crowdsourcing Partially-Subjective
  Ratings}.
\newblock In {\em {Proceedings of the fourth AAAI Conference on Human
  Computation and Crowdsourcing (HCOMP)}},  149--158.

\bibitem[\protect\citeauthoryear{Raykar and Yu}{2011}]{Rayker2011}
Raykar, V.~C., and Yu, S.
\newblock 2011.
\newblock Eliminating spammers and ranking annotators for crowdsourced labeling
  tasks.
\newblock {\em Journal of Machine Learning Research} 13:491--518.

\bibitem[\protect\citeauthoryear{Sheshadri and Lease}{2013}]{Sheshadri2013}
Sheshadri, A., and Lease, M.
\newblock 2013.
\newblock {SQUARE: A Benchmark for Research on computing crowd consensus}.
\newblock In {\em Proceedings of the AAAI conference on Human Computation
  (HCOMP)},  2035--2043.

\bibitem[\protect\citeauthoryear{Stuart \bgroup et al\mbox.\egroup
  }{2003}]{Stuart03}
Stuart, J.; Segal, E.; Koller, D.; and Kim, S.
\newblock 2003.
\newblock A gene-coexpression network for global discovery of conserved genetic
  modules.
\newblock {\em Science} 302(5643):249--255.

\bibitem[\protect\citeauthoryear{Whitehill \bgroup et al\mbox.\egroup
  }{2009}]{Whitehill2009}
Whitehill, J.; Ruvolo, P.; Wu, T.; Bergsma, J.; and Movellan, J.
\newblock 2009.
\newblock Whose vote should be count more: Optimal integration of labels from
  labelers of unknown expertise.
\newblock In {\em Proceedings of Advances in Neural Information Processing
  Systems},  2035--2043.

\end{thebibliography}
%\end{small}
\end{document}